\documentclass[aps,prl,reprint,superscriptaddress,showpacs]{revtex4-1}

\usepackage[]{hyperref} 
\usepackage[cp1250]{inputenc}
\usepackage{amsmath}
\usepackage{amsfonts}
\usepackage{textcomp}
\usepackage[]{graphicx}
\usepackage{epstopdf}
\usepackage{mathrsfs}
\usepackage{amsbsy}
\usepackage{cleveref}
\usepackage[rightcaption]{sidecap}

\graphicspath{ {./Figures-draft/} }

\crefformat{equation}{Eq.~(#2#1#3)}
\crefformat{figure}{Fig.~#2#1#3}

\Crefformat{equation}{Equation~(#2#1#3)}
\Crefformat{figure}{Figure~#2#1#3}

\DeclareMathOperator{\erf}{erf}

\DeclareMathOperator{\di}{d\kern-0.4ex}

\hypersetup{pdfauthor={Wiktor Walasik, Natalia M. Litchinitser},pdftitle={Multi-Filamentation}}

\begin{document}

\title{Dynamics of large femtosecond filament arrays: possibilities, limitations, and trade-offs}

\author{Wiktor Walasik}
\affiliation{Department of Electrical Engineering, University at Buffalo, The State University of New York, Buffalo, New York 14260, USA}
\author{Natalia M. Litchinitser}
\affiliation{Department of Electrical Engineering, University at Buffalo, The State University of New York, Buffalo, New York 14260, USA}
\email[]{natashal@buffalo.edu}

\date{\today}

\begin{abstract}
Stable propagation of large, multifilament arrays over long distances in air paves new ways for microwave-radiation manipulation. Although, the dynamics of a single or a few filaments was discussed in some of the previous studies, we show that the stability of large plasma filament arrays is significantly more complicated and is constrained by several trade-offs. Here, we analyze the stability properties of rectangular arrays as a function of four parameters: relative phase of the generating beams, number of filaments, separation between them, and initial power. We find that arrays with alternating phase of filaments are more stable than similar arrays with all beams in phase. Additionally, we show that increasing the size of an array increases its stability, and that a proper choice of the beam separation and the initial power has to be made in order to obtain a dense and  regular array of filaments. 
\end{abstract}


\maketitle

High-power femtosecond laser-pulse filamentation is a flourishing field of nonlinear optics due to its numerous applications in remote sensing, lightning protection, virtual antennas, and waveguiding~\cite{Fibich01,Couarion07,Chin12}.
Experiments show that a femtosecond laser beams can ionize atmosphere at the kilometer-scale distances~\cite{Durand13}. Propagation over such long distances is a result of the dynamic balance between the focusing Kerr nonlinearity, diffraction, and plasma defocusing due to multiphoton absorption.
A proper arrangement of plasma channels into arrays built of multiple filaments can lead to control~\cite{Marian13} and efficient guiding of electromagnetic radiation in air~\cite{Jhajj14}. Dielectric~\cite{Alshershby12a} and hollow-core metallic waveguide configurations~\cite{Musin07,Alshershby12b} were predicted theoretically and demonstrated experimentally~\cite{Chateauneuf08} to guide microwave radiation. Periodic arrays of densely packed high-intensity filaments were shown to create a hyperbolic metamaterial medium, that allows for radar signal manipulation and resolution enhancement~\cite{Kudyshev13,Will15}.

Formation of desired, regular filament patterns requires precise control of filament distribution and interaction. Multiple filaments can be generated by an intense Gaussian beam~\cite{Berge04,Fibich04}, whose power is much higher than the critical power of self-focusing $P_{\textrm{cr}}$. However, the distribution of filaments resulting from the breakup of a Gaussian beam is determined by intensity fluctuations~\cite{Mechain04} and modulation instability~\cite{Bespalov66,Zakharov09}. Moreover, the density of filaments generated in this way is limited~\cite{Henin10,Ettoumi15}. Therefore, this method can not provide the fine control and the high filament density required for the microwave-radiation manipulation. A certain degree of control of the filament distribution can be obtained by special preparation of the Gaussian beam. It was demonstrated experimentally that the predetermined initial phase~\cite{Liu11,Gao13}, amplitude distribution~\cite{Mechain04}, and geometry of the beam~\cite{Dubietis04} offer a limited control of the filamentation pattern. The filament distribution in an array can be even more deterministic if the position of each beam is managed independently. This is obtained by multiple beam generation using arrays of axicons~\cite{Sun12,Sun13} or microlenses~\cite{Camino13,Camino14,Xi15}. 

The question motivating our study is how to ensure that the initial multifilament distribution remains unchanged during the propagation or evolves in a predictable manner. This evolution is governed by the stability of a single filament in the array and by the mutual interaction between the neighboring filaments. A majority of studies of filament interaction focused only on two interacting beams and it was shown to result in attraction or repulsion of the filaments depending on their phase difference~\cite{Cai09,Ma08}, or rotation of the filament pattern~\cite{Shim10,Barbieri14}. Studies of two beams give a basic understanding of the filament interaction, but do not provide the insight on the interaction of beams in two-dimensional (2D) arrays build of multiple filaments. 

In this paper, we present a comprehensive study of the stability of large 2D arrays of filaments. We analyze the effects of the relative phase difference between the beams in the array on the propagation dynamics. Moreover, we study the effects related with the size of the array, the density of beams (separation distance) and the initial power.  In particular, we show that arrays built of out-of-phase filaments are more robust than these with all filaments in phase. Additionally, increase of the array size and a careful choice of the power and the beam separation allow us to increase the propagation distance over which the array of the beams propagates in a stable and predictable manner. This study provides guidance in choice of parameters in future experiments with large filament arrays.

This paper is organized in the following way. First we briefly discuss the general properties of structures that can be created using large arrays of plasma filaments in air, motivating our study on how to enable generation of stable arrays with required properties. Next, we describe methods used in our simulations and parameters of the studied system. Then, we investigate the difference in propagation dynamics of filament arrays with all the beams in phase and the arrays with alternating phase of the neighboring beams (out-of-phase arrays). Later, we consider the effects of the number of beams in the array, beam separation, and initial power for the out-of-phase arrays.

\begin{figure}[!t]
	\includegraphics[width = \columnwidth, clip=true, trim = {0 0 0 0}]{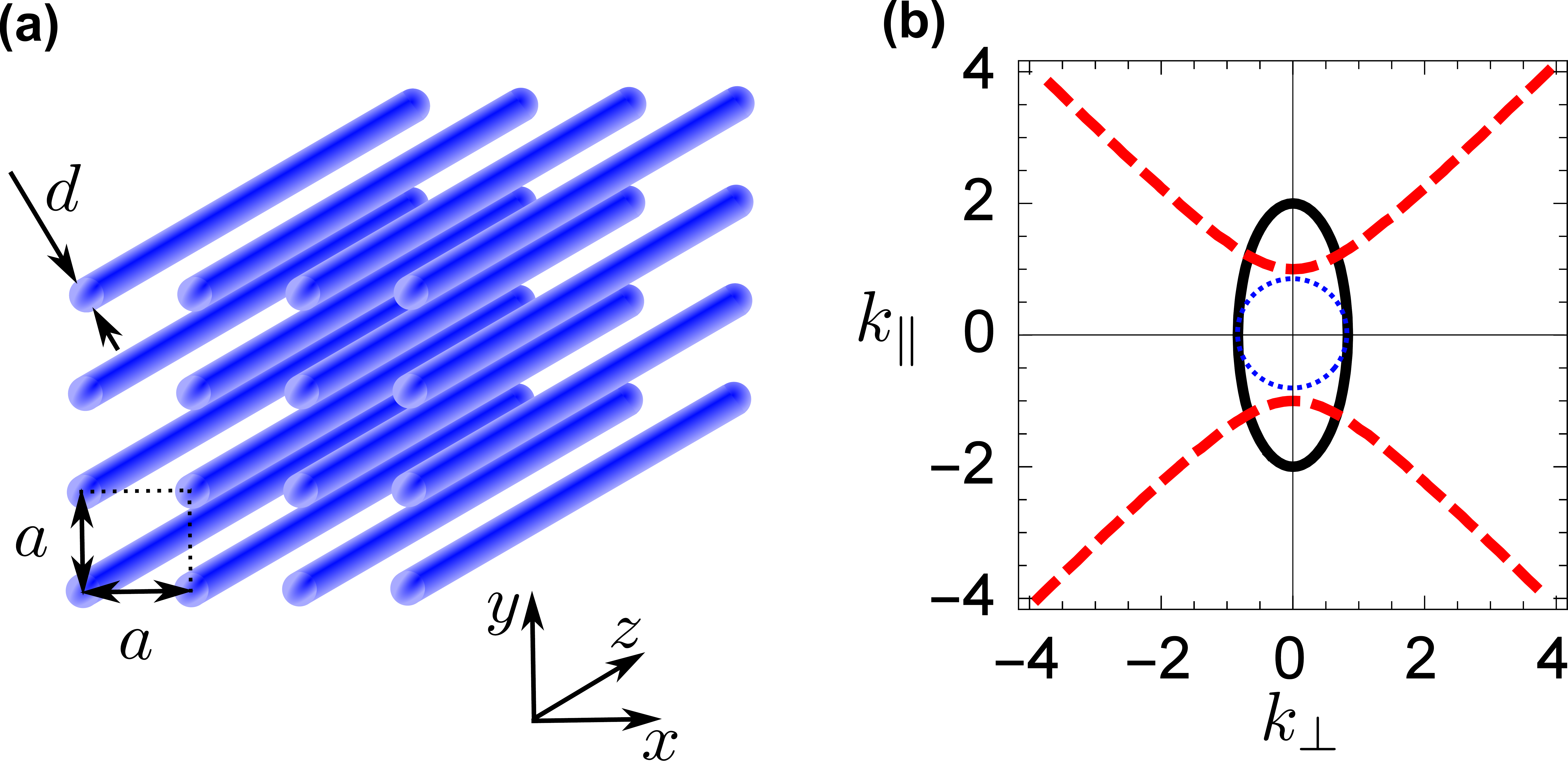}
	\caption{(a) Schematic illustration of the metamaterial formed by laser-induced plasma filaments (blue wires). The diameter of a filament is denoted as $d$, and $a$ represents separation between filaments. (b) Typical isofrequency contours for isotropic ($\epsilon_{\perp} = \epsilon_{\parallel}$---blue dotted curve), elliptical ($0 <\epsilon_{\parallel}<\epsilon_{\perp}$---black solid curve), and hyperbolic ($\epsilon_{\parallel} < 0 <\epsilon_{\perp}$---red dashed curve) medium.}
	\label{fig:hyper}
\end{figure}

Arrays of filaments, like the one presented in \cref{fig:hyper}(a), are able to create strongly anisotropic media, such as elliptical or hyperbolic medium. In the effective medium approximation, the diagonal elements of the permittivity tensor of such arrays are given by
\begin{subequations}
	\begin{align}
	\epsilon_{xx} = \epsilon_{yy} = \epsilon_{\perp} &= \epsilon_{h} \frac{(1+f)\epsilon_{\textrm{fil}} + (1-f)\epsilon_h}{(1-f)\epsilon_{\textrm{fil}} + (1+f)\epsilon_h},\\
	\epsilon_{zz} = \epsilon_{\parallel} &= f\epsilon_{\textrm{fil}} + (1-f)\epsilon_h,
	\end{align}
\end{subequations}
where $f = \pi[d/(2a)]^2$ denotes the filament filing fraction, $a$ represents the separation between the filaments in a square lattice, and $d$ is the filament diameter. Here, $\epsilon_h$ is the host medium permittivity and $\epsilon_{\textrm{fil}}$ is the permittivity of a filament. Filament permittivity can be described by the Drude model, in which the plasma frequency is related to the electron density in the plasma generated in a filament. The dispersion of extraordinary waves propagating in the $z$-direction in such a uniaxial medium is characterized by 
\begin{equation}
\frac{{k_{\parallel}^2}}{\epsilon_{\perp}} + \frac{k_{\perp}^2}{\epsilon_{\parallel}}  =\frac{{\omega^2}}{c^2}.
\label{eqn:iso}
\end{equation}
Typical solutions of \cref{eqn:iso} are shown \cref{fig:hyper}(b).

In order to study the nonlinear dynamics of filament propagation we use the (3+1)D nonlinear Schr\"odinger equation with the typical parameters for air~\cite{Couarion07,Skupin04}:
\begin{align}
\label{eqn:NLSE-general}
\frac{\partial E}{\partial z} = & \frac{i}{2 k_0} \left( \frac{\partial^2}{\partial x^2} + \frac{\partial^2 }{\partial y^2} \right) E 	- i \frac{k''}{2} \frac{\partial^2 E}{\partial t^2} +  \\
&i k_0 n_2 \mathcal{R}(t) E - \left( \frac{\sigma_B}{2} + i \frac{k_0}{2\rho_c} \right) \rho E - \frac{\beta^{(K)}}{2} |E|^{2K-2}E,  \nonumber
\end{align}
where the slowly varying electric field envelope $E$ is normalized in such a way that $|E|^2$ is the light intensity expressed in W/m$^2$. $k_0 = 2\pi/\lambda$ denotes the free-space wavevector and $\lambda = 775$~nm is the free-space wavelength. The group velocity dispersion is given by $k'' = 0.2$~fs$^2$/cm and the Kerr nonlinear parameter is $n_2 =5.57\cdot 10^{-19}$~cm$^2$/W. The inverse Bremsstrahlung cross-section responsible for plasma absorption is denoted by $\sigma_B = 5.54 \cdot 10^{-20}$~cm$^2$ and the critical plasma density at which air becomes transparent is given by $\rho_c = 1.86 \cdot 10^{-21}$~cm$^{-3}$. The multiphoton absorption coefficient is given by $\beta^{(K)} =3.1 \cdot 10^{-95}$~cm$^{13}$/W$^7$, where $K=8$ is the number of photons needed to overcome the ionization potential $U_i = 12$~eV. The time evolution of the electron density $\rho$ is governed by equation
\begin{equation}
\frac{\partial \rho}{\partial t} = \sigma_K \rho_{\textrm{nt}}|E|^{2K} + \frac{\sigma_B}{U_i}\rho |E|^2,
\label{eqn:rho-diff}
\end{equation}
where
$\sigma_K = \beta^{(K)}/( K \hbar \omega \rho_{\textrm{nt}})$ is the multiphoton ionization coefficient, $\hbar\omega$ denotes the energy of a single photon at the wavelength $\lambda = 2 \pi c/\omega$, and $\rho_{\textrm{nt}} = 5.4 \cdot10^{18}$~cm$^{-3}$ is the density of neutral oxygen molecules. Function
	\begin{equation}
	\mathcal{R}(t) = (1-\alpha) |E(t)|^2 + \frac{\alpha}{\tau_K} \int_{-\infty}^{t} e^{-(t-\tau)/\tau_K} |E(\tau)|^2 \textrm{d}\tau
	\label{eqn:NL-Raman}
	\end{equation}
models Kerr and non-instantaneous nonlinear responses~\cite{Skupin04}.
Here, $\alpha = 0.5$ is the fraction of the delayed nonlinear response due to stimulated molecular Raman scattering, and $\tau_K = 70$~fs denotes the characteristic relaxation time for oxygen molecules. For these parameters, the clamping intensity is $I_c = 31$~TW/cm$^2$.

Numerical solution of the full (3+1)D nonlinear Schr\"odinger equation is a very time and memory consuming process~\cite{Skupin04}. Therefore, we will simplify the problem using a reduced model proposed in Ref.~\cite{Skupin04}. This model has been successfully used in studies of multiple filamentation, for instance in Refs.~\cite{Huang13,Sun13,Shim10,Ma08}. Here we recapitulate the derivation of the reduced model along the lines presented in Ref.~\cite{Skupin04}.

We assume that the temporal distribution of the filament electric field can be described by a dominant Gaussian spike centered at a time instant~$t_c(z)$ with a time extent~$T$:  
\begin{equation}
E(x,y,z,t) = \psi(x,y,z) \exp\left[-\frac{[t-t_c(z)]^2}{T^2}\right].
\label{eqn:Gauss-temporal}
\end{equation}
Here $\psi(x,y,z)$ describes the spatial distribution of the electric field and $T = 0.1 t_c$, where $t_c=85$~fs is the duration of the pulse exciting the filament. The compression factor $0.1$ was shown in Ref.~\cite{Skupin04} to provide the best approximation of the full (3+1)D model. Under this assumption \cref{eqn:rho-diff} can be readily integrated to give
\begin{equation}
\rho = \frac{\sqrt{\pi}  T \sigma_K \rho_{\textrm{nt}} |\psi|^{2K} }{\sqrt{8K}}\left( \erf \left\{ \frac{\sqrt{2K} [t - t_c(z)] }{T} \right\} +1\right).
\label{eqn:rho-integrated}
\end{equation}
In the following derivation we neglect the dispersion term whose effects on the filament evolution are much smaller than these connected with ionization~\cite{Skupin04}. The plasma absorption is also neglected as it remains small compared to the multiphoton absorption. Inserting Eqs.~(\ref{eqn:Gauss-temporal})~and~(\ref{eqn:rho-integrated}) into \cref{eqn:NLSE-general} multiplied by $e^{-[t-t_c(z)]^2 / T^2}$ and integrating over the entire time domain yields the equation for~$\psi$:
\begin{align}
	\label{eqn:NLSE-simplified}
	\frac{\partial \psi}{\partial z} = & \frac{i}{2 k_0} \left( \frac{\partial^2}{\partial x^2} + \frac{\partial^2 }{\partial y^2} \right)  \psi  +  \\
	& i \theta k_0 n_2 |\psi|^2 \psi - i \gamma |\psi|^{2K} \psi - \frac{\beta^{(K)}}{2\sqrt{K}} |\psi|^{2K-2} \psi,  \nonumber
\end{align}
where the time-averaged nonlinear response is quantified by $\theta = [1 - \alpha + \alpha D/(\sqrt{2} \tau_K )]/\sqrt{2}$, the time-averaged loss coefficient is $	\gamma = (\sqrt{\pi/(8K)} T k_0 \sigma_K \rho_{\textrm{nt}} )/(2\rho_c)$, and  
	\begin{align}
	D = \int_{-\infty}^{+\infty} 
	& \textrm{exp}\left[\frac{T^2}{8 \tau_K^2} - \frac{u}{\tau_K} - 2 \left( \frac{u}{T}\right)^2 \right] \times \nonumber \\
	& \left[\erf\left(\frac{\sqrt{2}u}{T} - \frac{T}{\sqrt{8}\tau_K}\right) +1\right] \mathrm{d}u.
	\end{align}
The reduced model results in a 3D time-averaged nonlinear Schr\"odinger equation [\cref{eqn:NLSE-simplified}]. In order to study the evolution of the multifilament arrays we solve \cref{eqn:NLSE-simplified} using split-step Fourier method~\cite{Feit78,Lax81}.

The input field distribution $\psi(x,y,0)$ is described as a sum of Gaussian beams organized in a rectangular lattice:
\begin{equation}
 \psi(x,y,0) = A \sum\limits_{i=1}^{N_b} p^i \exp\left[ -\frac{(x-x_{0,i})^2 + (y-y_{0,i})^2}{\sigma^2} \right],
 \label{eqn:input}
\end{equation}
where $N_b$ denotes the number of beams and $p^i$ is the phase of the $i$th beam. In an in-phase array, all the beams have the same phase and $p=1$. On the contrary, in an out-of-phase array, each beam has the phase opposite to its nearest neighbors and therefore (for rectangular arrays with odd number of beams studied here) $p=-1$. $\sigma$~is the $1/e$ beam width and a pair [$x_{0,i}, y_{0,i}$] denotes the location of the center of the $i$th beam in the transverse plane. The field amplitude $A$ is related to the initial power. The initial power per beam normalized to the critical power in air $P_{\textrm{cr}} = 3.77\lambda^2/(8\pi n_2) = 1.6$~GW is calculated as 
\begin{equation}
P_{\textrm{in}} = \frac{1}{P_{\textrm{cr}} N_b}\int_{-\infty}^{\infty} \int_{-\infty}^{\infty}  \psi(x,y,0) \mathrm{d}x \mathrm{d}y.
\label{eqn:power}
\end{equation}
In our studies, input beams have spatial width $\sigma = 0.15$--$0.20$~mm, which is approximately twice the width of a filament. Intensity distributions similar to our inputs were generated by a wide Gaussian beam impinging on an array of microlenses~\cite{Camino13,Camino14,Xi15}.

\begin{figure}[!t]
	\includegraphics[width = 0.49\columnwidth, clip=true, trim = {0 0 30 0}]{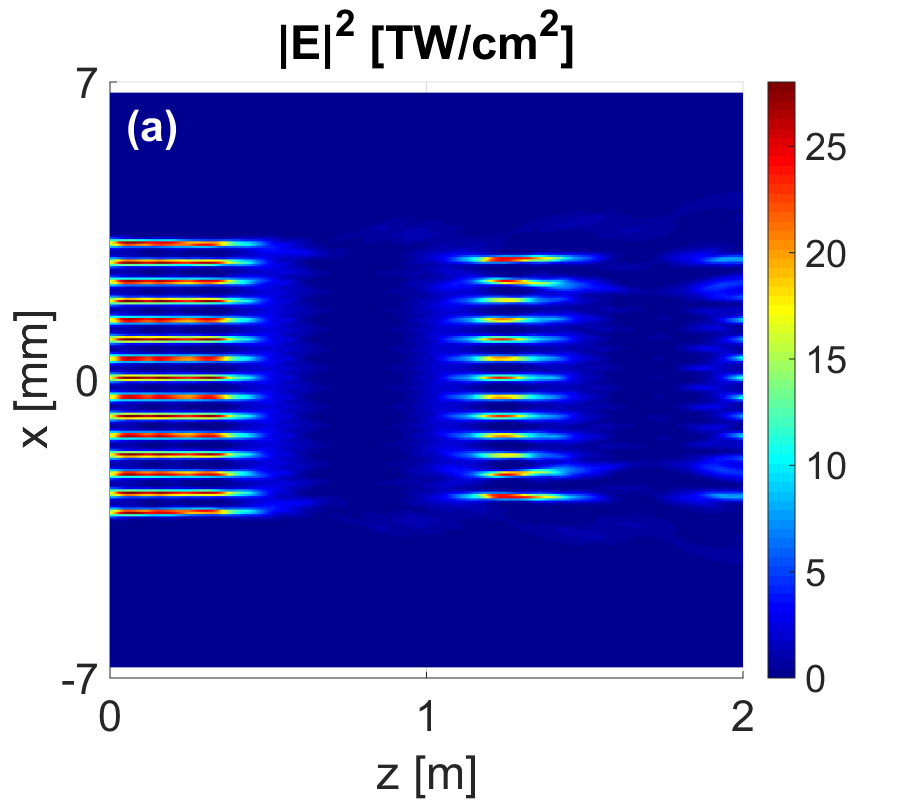}
	\includegraphics[width = \columnwidth, clip=true, trim = {30 0 80 0}]{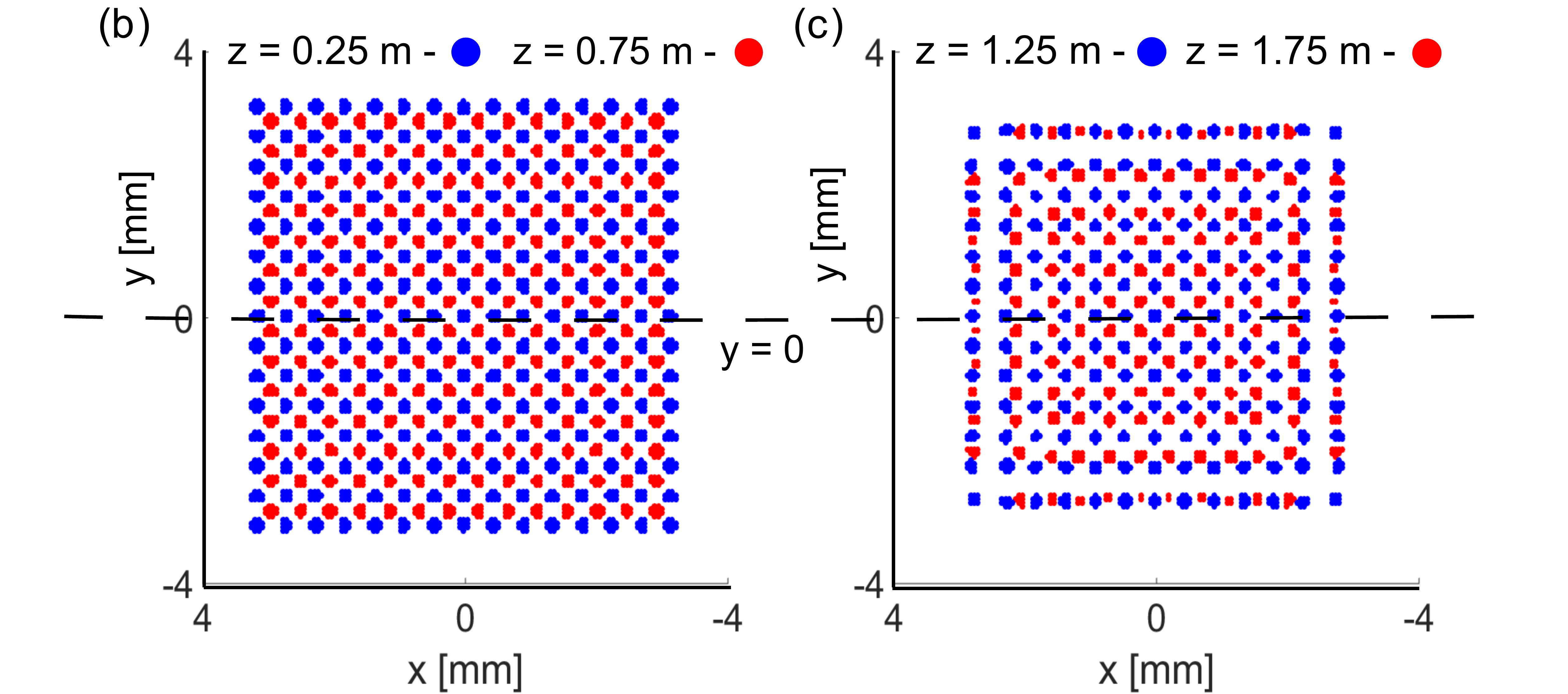}
	\caption{(a) Filament intensity distribution $|E(x,y=0,z)|^2$ for the in-phase $15 \times 15$ beam array. (b)--(c) Superimposed filament location in the transverse plane at four different propagation distances: (b) $z = 0.25$~m (blue) and $z=0.75$~m (red); (c) $z = 1.25$~m (blue) and $z=1.75$~m (red). We assume that the filament is formed when the light intensity is higher than~$I_c/8$. Array parameters are: the normalized initial power per beam~$P_{\textrm{in}} = 4$, the width of a single beam $\sigma = 0.15$~mm, and the separation between beams $a = 0.45$~mm.}
	\label{fig:beam-location}
\end{figure}

Let us first consider a large ($15 \times 15$) in-phase array of Gaussian beams forming filaments. \Cref{fig:beam-location}(a) shows the light intensity distribution $E(x,z)$ in the plane $y = 0$, whereas Figs.~\ref{fig:beam-location}(b) and (c) show the location of the beams in the transverse plane ($x$--$y$) for different values of the propagation distance $z$. At the beginning of the propagation ($z= 0$--$0.5$~m), a regular $15 \times 15$ array of filaments is formed and the central beams lay in the plane $y=0$. Because all the beams in the array have the same phase their mutual interaction results in attraction of the neighboring beams and the beams fuse at $z \approx 0.5$~m. This results in a smaller array of $14 \times 14$ beams shown in~\cref{fig:beam-location}(b) in red. None of the beams in this array is located in $y=0$ plane. This is the reason of the apparent light intensity decrease in the map shown in~\cref{fig:beam-location}(a). Similar fusion process leads to the generation of a $13 \times 13$ array [shown in blue in~\cref{fig:beam-location}(c)] visible in~\cref{fig:beam-location}(a) and a $12 \times 12$ array [shown in red in~\cref{fig:beam-location}(c)] not visible in~\cref{fig:beam-location}(a). As a result of mutual filament attraction, the array of the in-phase beams decreases in size and loses its regularity. The outer beams increase their distance from the center of the array, dissipate the energy, and are unable to reform filaments which accelerates shrinking process of the array.

\begin{figure}[!t]
	\includegraphics[width = 0.49\columnwidth, clip=true, trim = {0 -80 -120 0}]{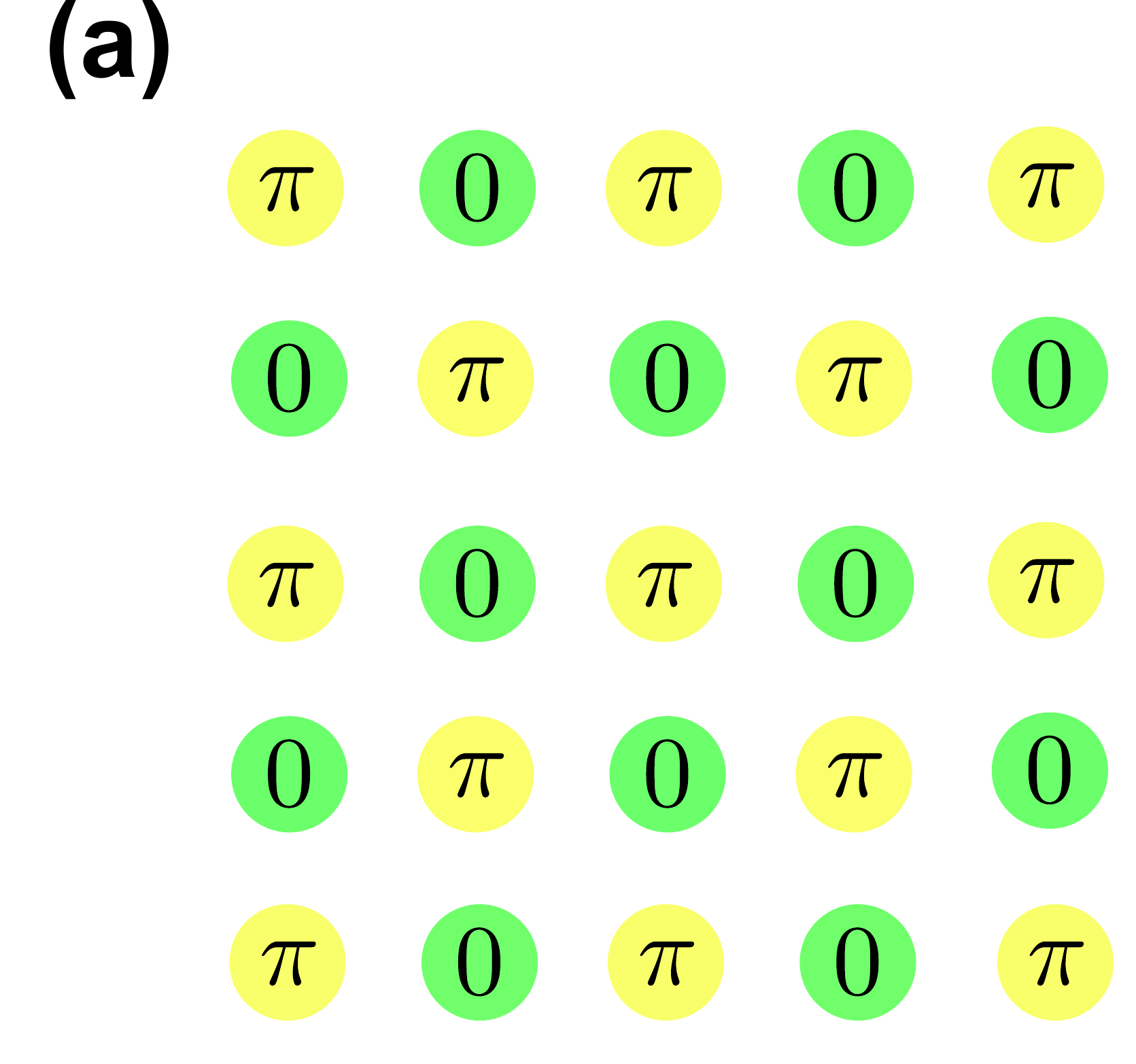}
	\includegraphics[width = 0.49\columnwidth, clip=true, trim = {0 0 30 0}]{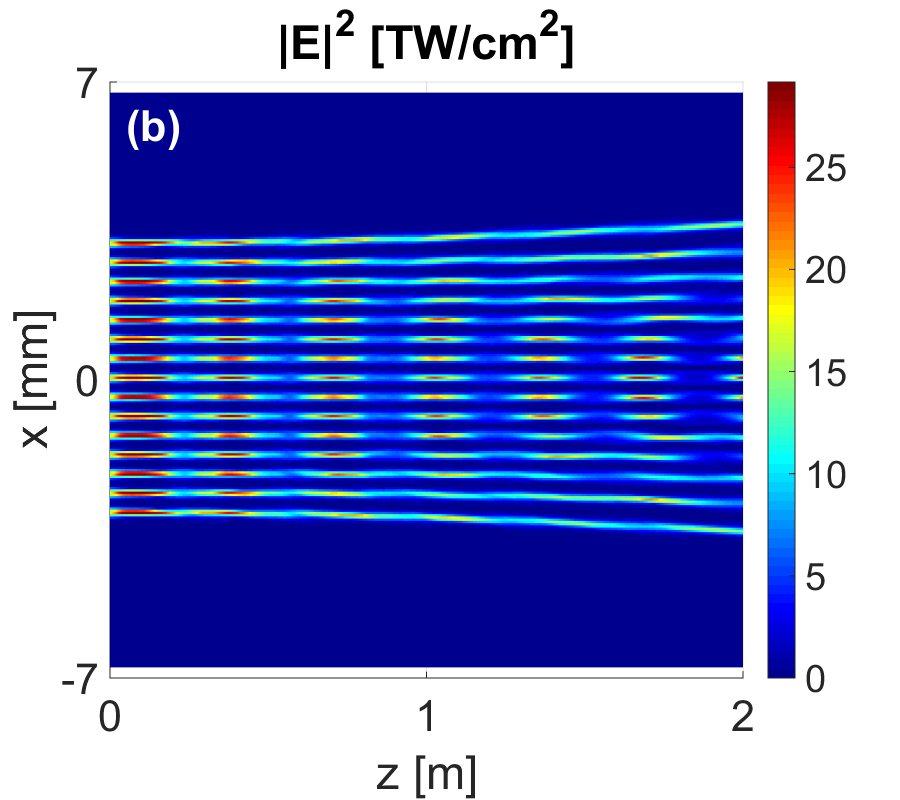}
	\includegraphics[width = 0.49\columnwidth, clip=true, trim = {0 0 0 0}]{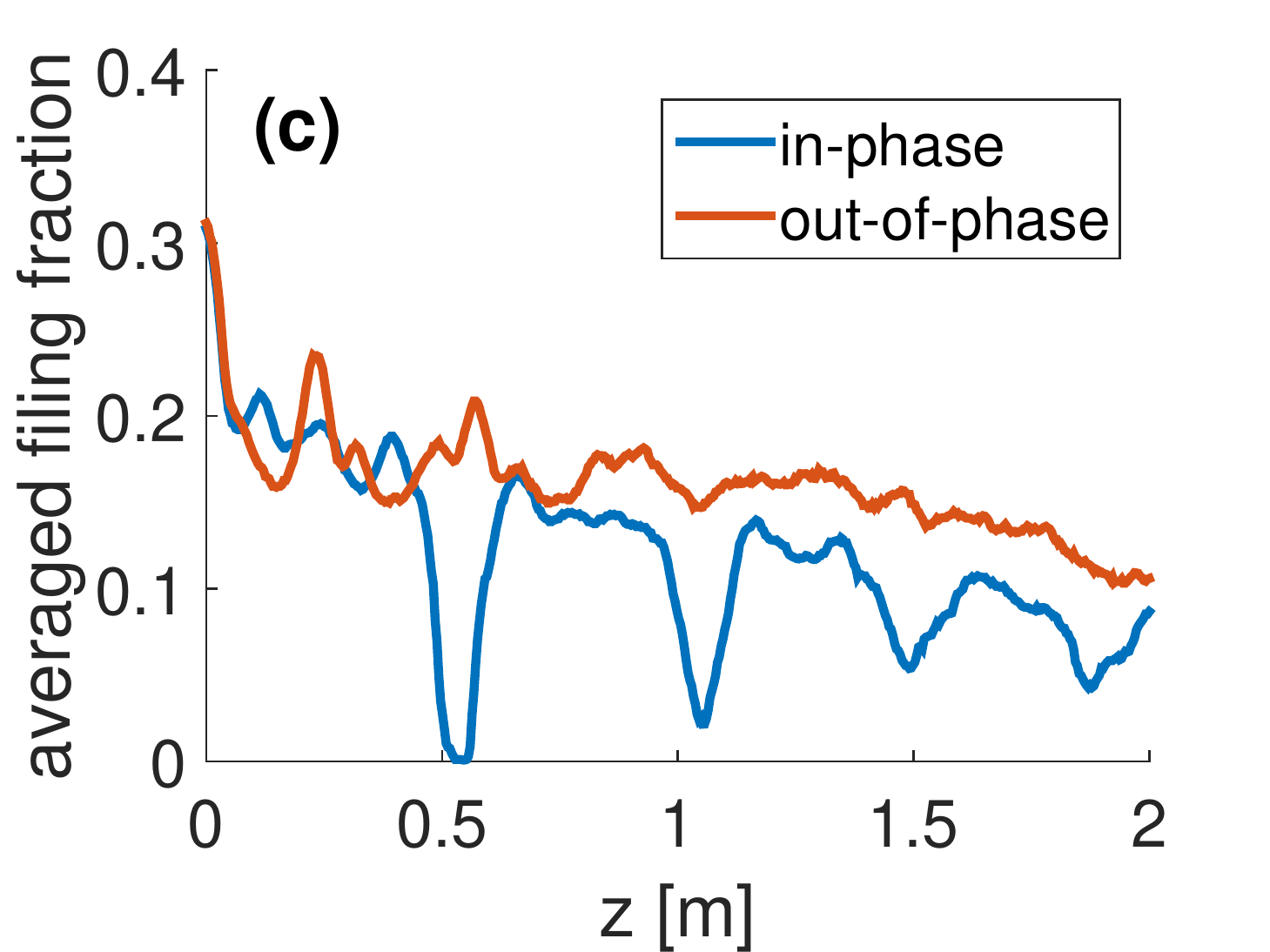}
	\caption{(a) Chessboard-like phase distribution in the out-of-phase filament array. (b)  Filament intensity distribution $|E(x,y=0,z)|^2$ for the out-of-phase $15 \times 15$ beam array. Other parameters of the array are identical to these in \cref{fig:beam-location}. (c) Comparison of the averaged filament filing fraction $\langle f \rangle$ as a function of propagation distance $z$ for the out-of-phase $15 \times 15$ array and the in-phase presented in \cref{fig:beam-location}.}
	\label{fig:oof}
\end{figure}

In order to compare the in-phase and the out-of-phase arrays we study an array with the same parameters as those used in \cref{fig:beam-location} but with the alternate phase of the beams. Such a phase distribution is schematically presented in \cref{fig:oof}(a). The nearest neighbors of each of the beams are out-of-phase with the center beam, forming a chessboard-like pattern. \Cref{fig:oof}(b) shows the light intensity distribution $E(x,z)$ in the plane $y = 0$ for the out-of-phase array of beams. This time, beams do not merge and the number of filaments in the array is conserved during the propagation. This is the result of the mutual repulsion between the nearest-neighbor out-of-phase beams. The peak intensity of light in the filament beams oscillates during the propagation and reflects the dynamical balance between the Kerr focusing and the plasma defocusing. As a result of their mutual repulsion, the outer beams increase their distance from the center of the array even faster than in the case of the in-phase array. Nevertheless, the out-of-phase array offers a significantly more stable behavior than the in-phase array since the number of filaments and the position of beams far from the edges of the array do not change. The stability of the two types of arrays can be also compared using the averaged filing fraction of filaments $\langle f \rangle$ shown in \cref{fig:oof}(c). The averaged filing fraction is computed as a ratio between the transverse area where the filaments are present and the area of the whole array (the area of the array may increase during the propagation due to the divergence of beams located close to the edges of the array). The presence of a filament is determined by the fact that the light intensity at a given position is higher than a certain threshold value. In this study we chose this threshold to be $1/8$ of the clamping intensity $I_c$. For the in-phase array, the filing fraction decreases and experiences abrupt changes when array changes its structure (when filaments merge). The decrease of the filing fraction of the out-of-phase array is much more uniform. In the later stages of propagation ($z =1-2$~m), the filing fraction for the out-of-phase array is on average 1.5 times higher than for the in-phase array. These results suggest that the out-of-phase arrays are generally more stable than the in-phase arrays and therefore, further discussion will focus on the out-of-phase arrays. 

\begin{figure}[!t]
	\includegraphics[width = 0.46\columnwidth, clip=true, trim = {0 40 100 0}]{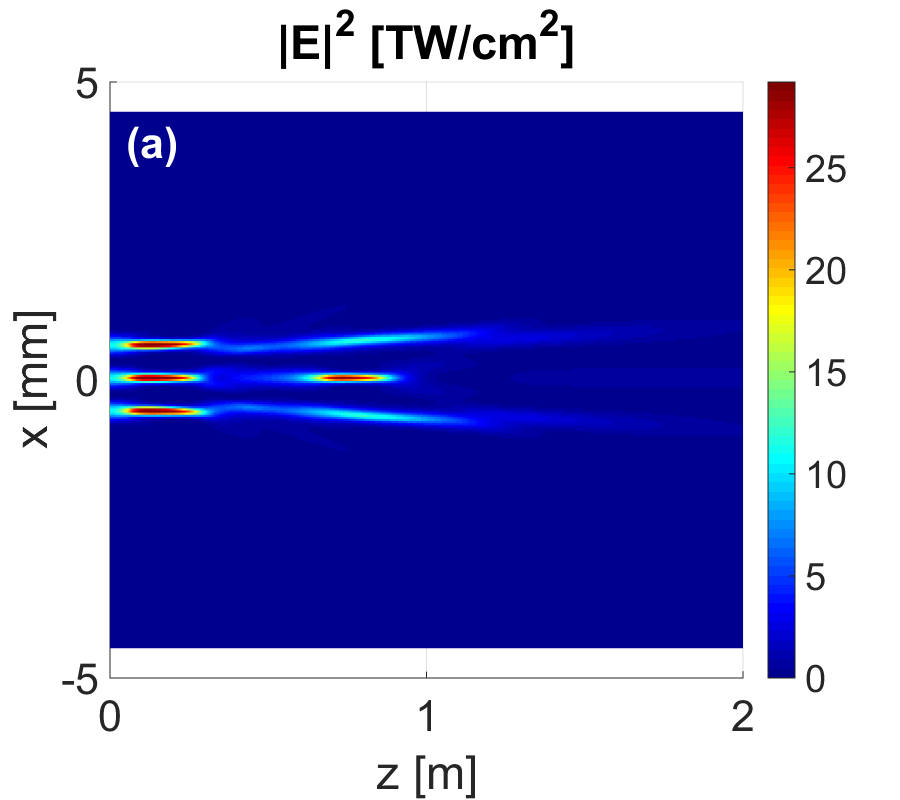}
	\includegraphics[width = 0.51\columnwidth, clip=true, trim = {43 40 0 0}]{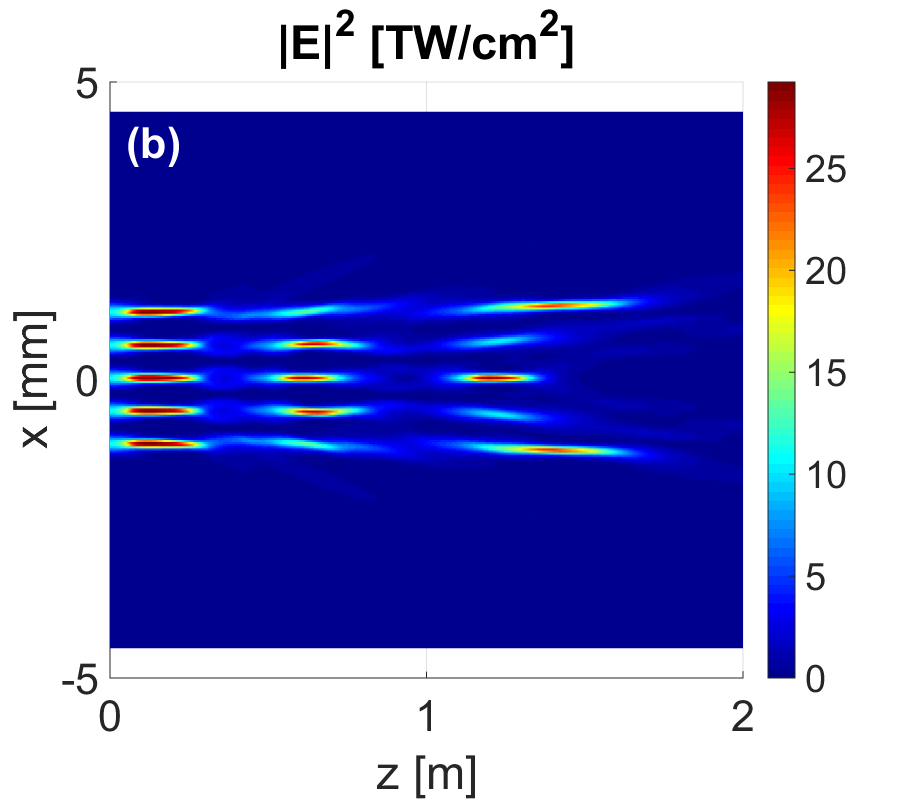}
	\includegraphics[width = 0.46\columnwidth, clip=true, trim = {0 0 100 0}]{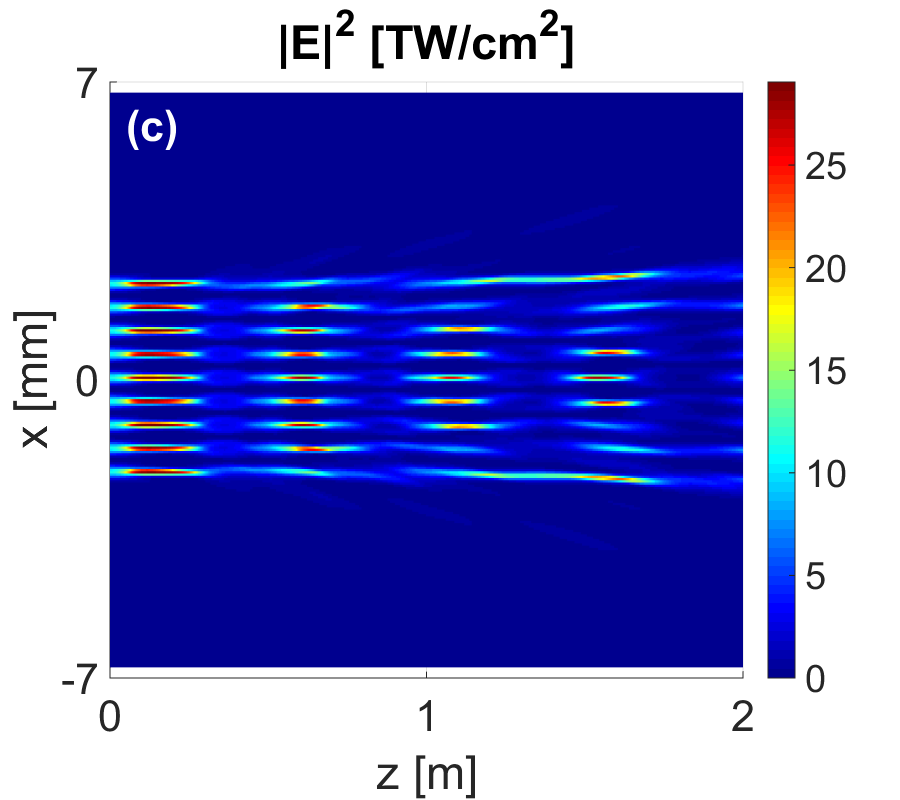}
	\includegraphics[width = 0.51\columnwidth, clip=true, trim = {43 0 0 0}]{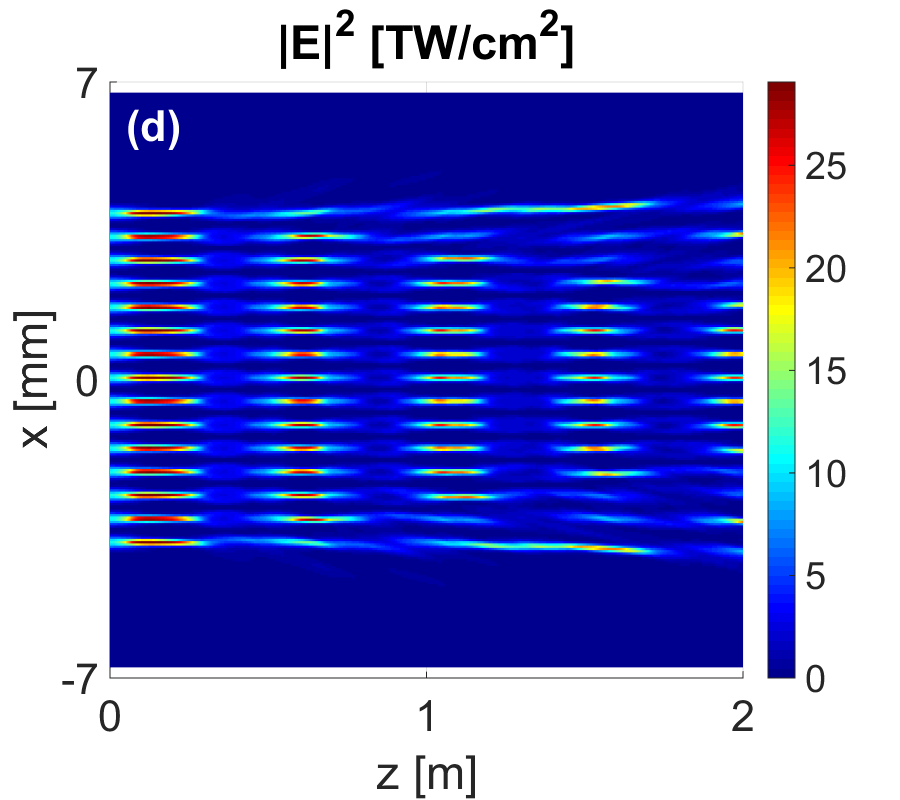}
	\caption{Light intensity distribution in the $y = 0$ plane for filament arrays of different sizes (number of beams $N_b$): (a) $3 \times 3 = 9$, (b) $5 \times 5 = 25$, (c) $9 \times 9  = 81$, and (d) $15 \times 15 = 225$. In each array $P_{\textrm{in}} = 4$, the beam separation is $a=0.55$~mm, and the width of a single beam $\sigma = 0.2$~mm. The $x$ range is different for plots (a), (b) and (c), (d).}
	\label{fig:nb-beams}
\end{figure}

Let us analyze the stability of the out-of-phase filament array as a function of the number of beams building it. \Cref{fig:nb-beams} shows the dynamics of arrays with different sizes. We show that the larger the array, the more stable it is. Small arrays, built of 9 or 25 beams, tend to destabilize faster than larger arrays. We notice that the outer filaments increase their distance from the center of the array, and then lose the energy due to diffraction. This effect can be explained in the following way. The most stable filaments are those in the middle of an array since they have all four (out-of-phase) nearest neighbors. Each of these neighbors repulses the central filament, but the net force is zero due to the symmetry reasons. The filaments on the edges and corners are more susceptible to transverse displacement as they only have three or two nearest neighbors, and the net repulsive force acting on them is non-zero. The larger the array, the higher is the ratio between the bulk and edge filaments, and consequently, the better is the stability of the filament array. The fact that the stability of the arrays increases with their size is a very desirable property from the experimental point of view, as the manipulation of microwave beams requires large beam arrays that are uniform or evolve in a predictable manner.

%

\begin{figure}[!t]
	\includegraphics[width = 0.46\columnwidth, clip=true, trim = {0 40 100 0}]{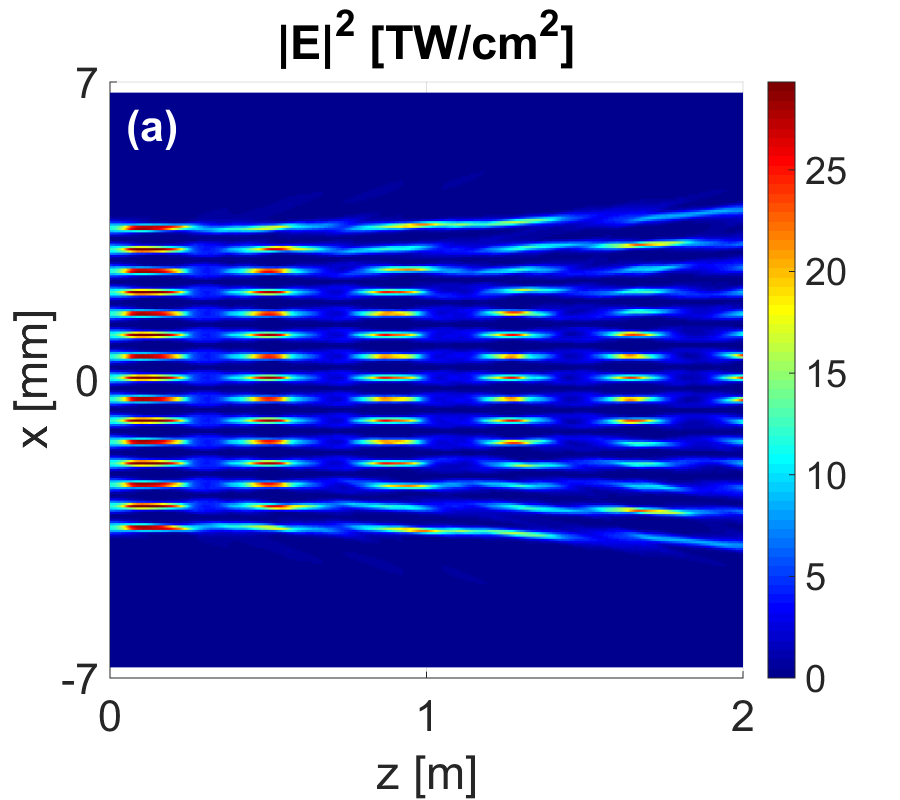}
	\includegraphics[width = 0.51\columnwidth, clip=true, trim = {43 40 0 0}]{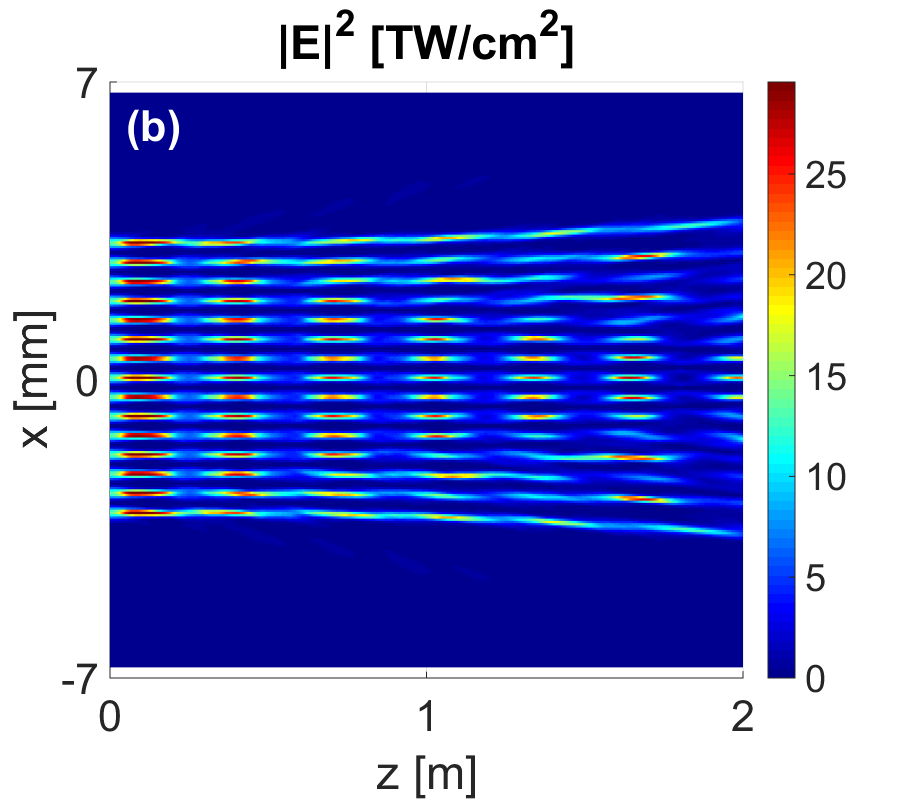}
	\includegraphics[width = 0.46\columnwidth, clip=true, trim = {0 0 100 0}]{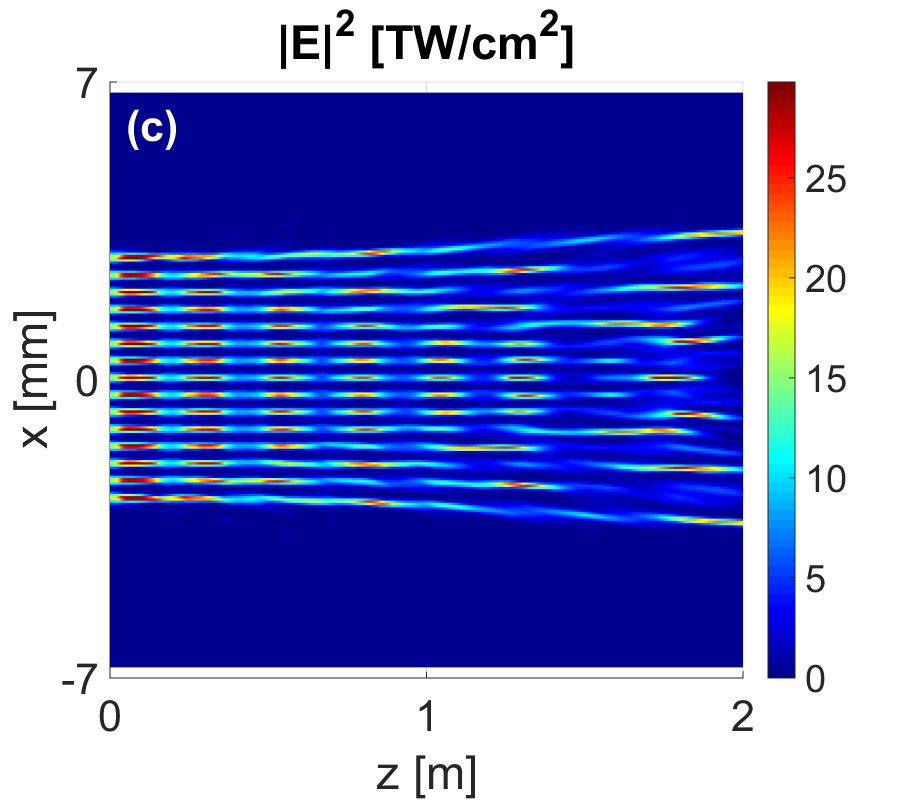}
	\includegraphics[width = 0.51\columnwidth, clip=true, trim = {43 0 0 0}]{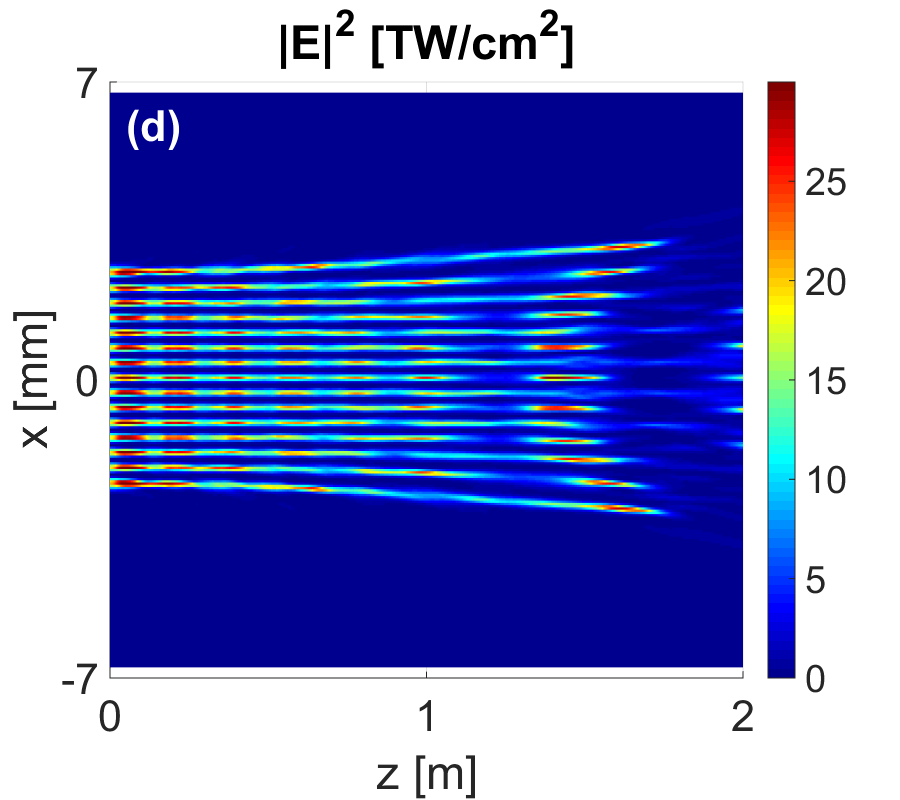}
	\includegraphics[width = 0.49\columnwidth, clip=true, trim = {0 0 0 0}]{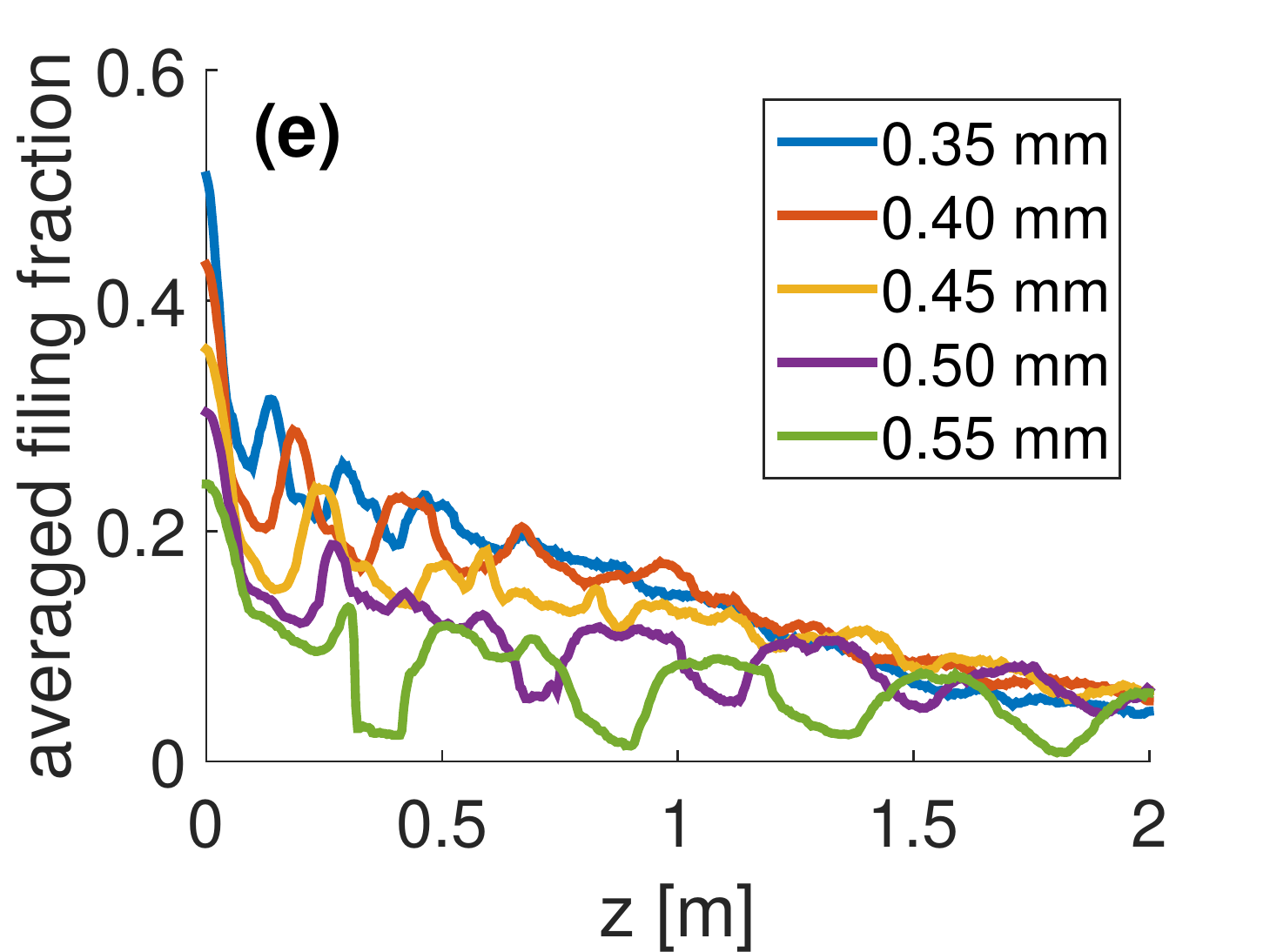}
	\caption{Light intensity distribution in the $y = 0$ plane for large ($15 \times 15$) filament arrays with different beam separation~$a$: (a) $0.50$~mm, (b) $0.45$~mm, (c) $0.40$~mm, and (d) $0.35$~mm. The evolution of an array with $a=0.55$~mm is presented in \cref{fig:nb-beams}(d). In each array $P_{\textrm{in}} = 4$ and the width of a single beam $\sigma = 0.2$~mm. (e) Comparison of the averaged filament filing fraction $\langle f \rangle$ as a function of propagation distance $z$ for the out-of-phase $15 \times 15$ arrays with different initial beam separation.}
	\label{fig:15-beams-OOP-sep}
\end{figure}

Next, we will study the behavior of the beam array as a function of the separation between the neighboring filaments. The separation of the beams determines the filing fraction of filaments, which is a critical parameter for the properties of filament-based waveguides and metamaterials~\cite{Kudyshev13}. From the practical viewpoint, high filing fractions are necessary for efficient microwave beam guiding and manipulation. However, as \cref{fig:15-beams-OOP-sep} shows, the increase to the filing fraction decreases the stability of the filament array. As a consequence of a decreased beam separation, the filament-filament interaction becomes stronger, and the array broadens. This effectively decreases the filing fraction as it can be seen from \cref{fig:15-beams-OOP-sep}(e). The arrays with the smallest beam separations have the highest filing fraction in the first 1~m of propagation. However, due to the rapid broadening of these arrays, the filing fraction decreases and at $z = 2$~m it is equal to the values obtained for arrays with larger beam separation. Therefore, the trade-off between the need for a high filing fraction and the array-size evolution should be taken into account.
The oscillations of the peak intensity of filaments are also reflected in the filing fraction. They are especially visible for arrays with larger beam separation, for which the filing fraction oscillates during the propagation. This result shows that, for arrays with $P_{\textrm{in}} = 4$, the filling fraction of densely packed arrays behaves more monotonically that for sparse arrays but experiences a larger overall decrease during the entire propagation distance studied here.


\begin{figure}[!t]
	\includegraphics[width = 0.505\columnwidth, clip=true, trim = {0 40 0 0}]{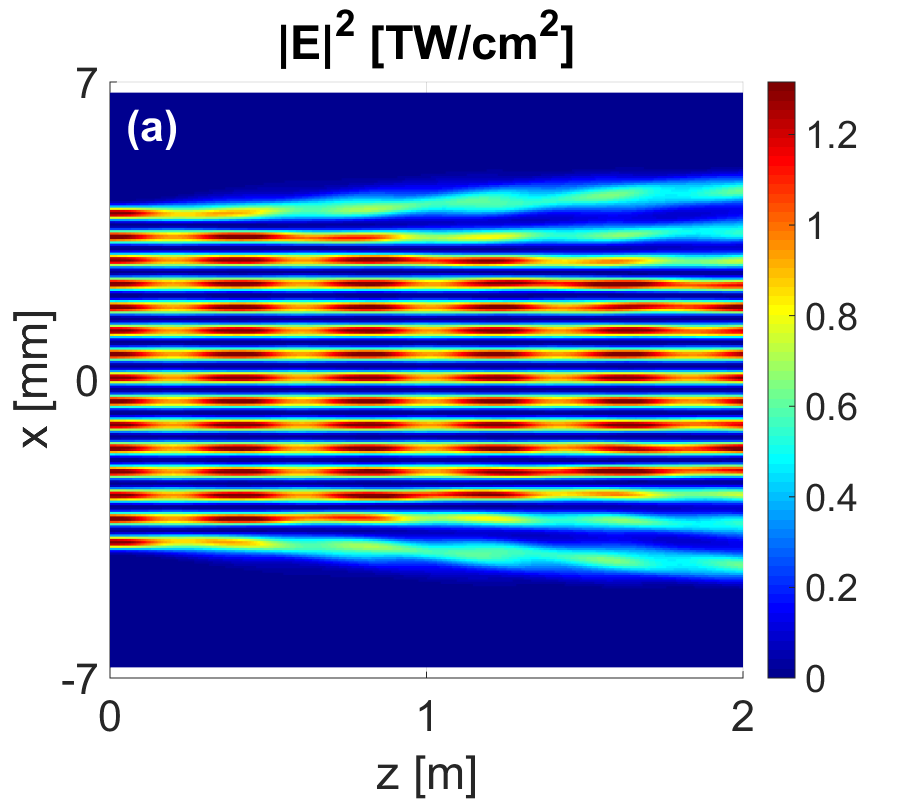}
	\includegraphics[width = 0.48\columnwidth, clip=true, trim = {43 40 0 0}]{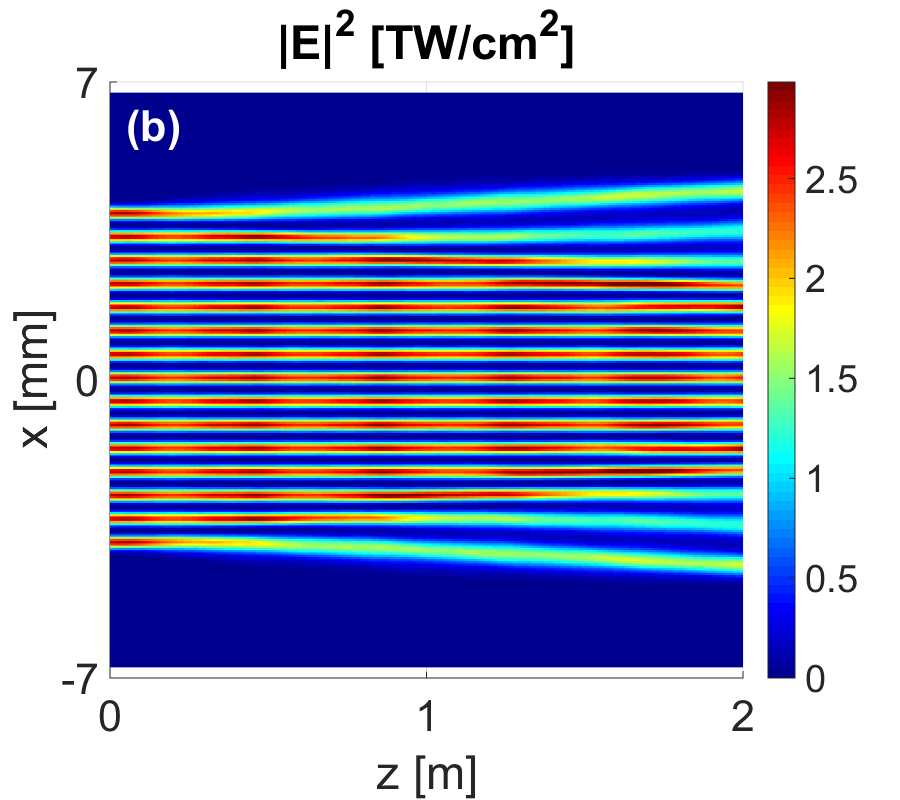}
	\includegraphics[width = 0.505\columnwidth, clip=true, trim = {0 0 0 0}]{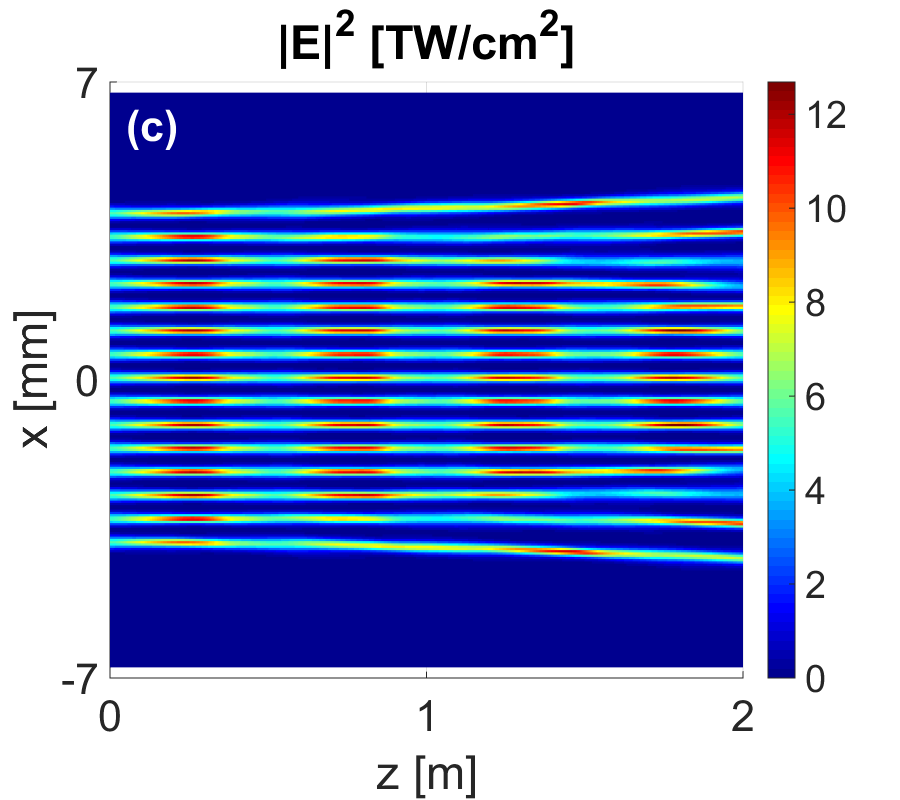}
	\includegraphics[width = 0.48\columnwidth, clip=true, trim = {43 0 0 0}]{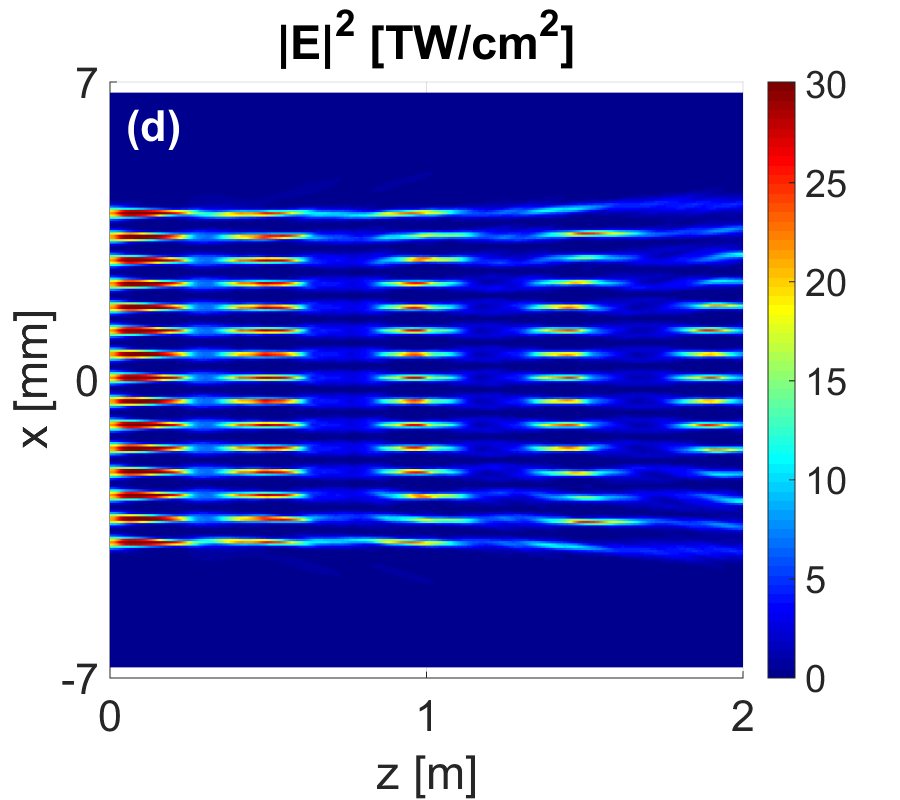}
	\caption{Light intensity distribution in the $y = 0$ plane for large ($15 \times 15$) filament arrays with different values of initial power $P_{\textrm{in}}$: (a) $0.5$, (b) $1$, (c) $2$, and (d) $8$. In each array the separation between beams is equal to $a=0.55$~mm and the width of a single beam $\sigma = 0.2$~mm. The evolution of an array with initial power of $P_{\textrm{in}} = 4$ is presented in \cref{fig:nb-beams}(d).}
	\label{fig:15-beamsOOP-pow}
\end{figure}

Finally, we studied the behavior of the filament array as a function of the initial power used to generate the array. As shown if \cref{fig:15-beamsOOP-pow}, at low powers ($P_{\textrm{in}} = 0.5$ or~$1$), the outer beams diverge away from the array center in early stages of propagation and diffract rapidly as the power is too low to create a filament. For a chosen beam width $\sigma = 0.2$~mm, the most stable propagation of a single beam is obtained for $P_{\textrm{in}} = 1$ [see \cref{fig:15-beamsOOP-pow}(b)]. It results in a practically constant peak intensity of beams close to the array center. However, the peak intensity obtained at $P_{\textrm{in}} = 1$ is ten times lower than the saturation intensity of a filament reached for arrays excited with high powers ($P_{\textrm{in}} = 4$ or~$8$). In high-power regime the divergence and diffraction of the outer filaments is slower than for low-power arrays but the oscillations of the peak intensity have higher amplitudes. These results suggest another trade-off between the uniformity of a single filament and the array as a whole. A power level resulting in a proper balance between the regularity of the whole array and the uniformity of a single filament should be chosen in experimental studies.

In conclusion, inspired by the possibility of designing large arrays of filaments for microwave manipulation, we have systematically studied the propagation of large arrays of femtosecond laser beams in air. We have analyzed the effect of the relative phase difference between the filaments in the array and found out that the arrays with a chessboard-like out-of-phase beam arrangement propagate unchanged for longer distances than the in-phase arrays. The in-phase arrays tend to reduce their size during the propagation due to mutual attraction and resulting fusion of the filaments. Furthermore, for the out-of-phase arrays we have analyzed the dependence of their stability on the following parameters: the number of beams building an array, the separation between the beams, and the initial power per beam. Our studies have shown that the increase of the number of filaments leads to longer stable propagation of an array. Similar stability enhancement of the array as a whole was observed with the increase of the excitation power, however, the oscillations of intensity inside a single filament increase at the same time. Moreover, we have shown that the reduction of the separation between the filaments has two effects: on the one hand it increases the filament filing fraction and reduces its oscillations during the propagation, on the other hand it accelerates the increase of the array transverse size. All the effects discussed above influence the attainable filament filing fraction and have a crucial importance for experimental realization of filament-based metamaterials.

\section{Acknowledgments}

This work was supported by US Army Research Office grants: \#W911NF-15-1-0146 and MURI \#W911NF-11-0297.


%

\section{Author contributions}
Both authors contributed equally to the study, discussed the results and commented on the manuscript at all stages. N.M.L. proposed the study of stability of large filament arrays. W.W. developed the codes, performed the parameter optimization, and numerical simulations. N.M.L. supervised the work.

\section{Additional Information}
\subsection{Competing financial interests}
The authors declare no competing financial interests.

\end{document}